\documentclass[12pt]{article}

\usepackage[body={17.5cm, 22.5cm},right=2cm]{geometry}

\usepackage{float}
\usepackage[mathscr]{euscript}

\usepackage{booktabs}
\usepackage{caption}

\usepackage{color}
\usepackage{graphicx}
\usepackage{epsf}
\usepackage{graphicx,epsfig}
\usepackage{amsmath}

\DeclareMathOperator{\arccosh}{arccosh}
 
\usepackage{multirow}

\usepackage{amsmath, amsthm, amssymb}

\usepackage{epsfig}
\usepackage{cite}
\usepackage{color,colordvi}

\newcounter{hran}

\makeatletter
\renewcommand\section{\@startsection {section}{1}{\z@}%
                               {-3.5ex \@plus -1ex \@minus -.2ex}%
                               {2.3ex \@plus.2ex}%
                               {\normalfont\large\bfseries}}
\makeatother

\begin{document}\thispagestyle{empty}

\vspace{0.5cm}

\def\thefootnote{\arabic{footnote}}
\setcounter{footnote}{0}

\def\s{\sigma}
\def\nn{\nonumber}
\def\p{\partial}
\def\ls{\left[}
\def\rs{\right]}
\def\lc{\left\{}
\def\rc{\right\}}
\def\oT{\overline{T}}
\def\oX{\overline{X}}
\def\oQ{\overline{Q}}
\def\oB{\overline{B}}
\newcommand{\oF}{\overline{F}}
\def\d{{\rm d}}
\newcommand{\be}{\begin{eqnarray}}
\newcommand{\ee}{\end{eqnarray}}
\newcommand{\bi}{\begin{itemize}}
\newcommand{\ei}{\end{itemize}}
\renewcommand{\th}{\theta}
\newcommand{\bth}{\overline{\theta}}
\def\simlt{\stackrel{<}{{}_\sim}}
\def\simgt{\stackrel{>}{{}_\sim}}
\vspace{0.5cm}
\hspace*{12cm}
CERN-PH-TH-2014-085

\vspace*{1cm}
\begin{center}

{\large \bf 
The Imaginary Starobinsky Model and Higher Curvature Corrections
}
\\[1.5cm]
{\large   Sergio Ferrara$^{a,b,c}$, Alex Kehagias$^{d,e}$ and Antonio Riotto$^{e}$}
\\[0.5cm]

\vspace{.3cm}
{\normalsize {\it  $^{a}$  Physics Department, Theory Unit, CERN,
CH 1211, Geneva 23, Switzerland}}\\

 \vspace{.3cm}
{\normalsize {\it  $^{b}$  INFN - Laboratori Nazionali di Frascati \\
Via Enrico Fermi 40, I-00044 Frascati, Italy}}\\

 \vspace{.3cm}
{\normalsize {\it  $^{c}$  Department of Physics and Astronomy,
University of California \\ Los Angeles, CA 90095-1547, USA}}\\

\vspace{.3cm}
{\normalsize {\it  $^{d}$ Physics Division, National Technical University of Athens, \\15780 Zografou Campus, Athens, Greece}}\\

\vspace{.3cm}
{\normalsize { \it $^{e}$ Department of Theoretical Physics and Center for Astroparticle Physics (CAP)\\ 24 quai E. Ansermet, CH-1211 Geneva 4, Switzerland}}\\

\vspace{.3cm}

\end{center}

\vspace{3cm}

\hrule \vspace{0.3cm}
{\small  \noindent \textbf{Abstract} \\[0.3cm]
\noindent 
We elaborate on  the predictions of the imaginary Starobinsky model of inflation coupled to matter, where the inflaton is 
identified with the imaginary part of the inflaton multiplet suggested by the Supergravity embedding of a pure $R+R^2$ gravity.
In particular, we study the impact of  higher-order curvature terms and  show that, depending on the parameter range, one may find either a quadratic model of chaotic inflation or  monomial models of chaotic inflation with fractional powers between 1 and 2.
\vspace{0.5cm}  \hrule

\vskip 1cm

\def\thefootnote{\arabic{footnote}}
\setcounter{footnote}{0}



\baselineskip= 19pt
\newpage 

\section{Introduction}
The recent Planck results \cite{Ade:2013uln} have indicated that the cosmological perturbations in the Cosmic Microwave
Background (CMB) radiation are nearly gaussian and of the adiabatic type. If it is assumed that 
 these scalar perturbations are generated by an inflationary slow-roll single-field \cite{lr}, the data put severe
restrictions on the inflationary  parameters. In particular, the Planck results have strengthened the upper
limits on the tensor-to-scalar ratio, $r< 0.12 $ at 95\% C.L., disfavouring many inflationary models.
Among the inflationary models still consistent with the Planck data is the $(R+R^2)$ Starobinsky model \cite{star} which predicts 
\be
\label{pred}
r\approx 12/N^2, 
\ee 
where $N={\cal O}(50)$ is the number of e-folds till the end of inflation. 
This  triggered a renewal interest in the Starobinsky model of inflation and  in particular in  its supersymmetric extentions,
first originated in  Refs. \cite{fv,Cecotti:1987sa,CFPS}  and recently developed in  
Refs. \cite{KL,sugrastaro,ENO,FKLP,FKR,FKD,KT,FKRim,Chakra}.

Although the Starobinsky model is  favoured by Planck data, it seems to be  disfavoured by the  recent BICEP2 data on the  large angle CMB B-mode polarization   \cite{BICEP} which indicate a  tensor-to-scalar ratio 
\be 
r=0.2^{+0.07}_{-0.05}.
\ee 
Indeed, in the supersymmetric embedding of the Starobinsky model the inflaton is a part of a chiral mutiplet $T$, the inflaton multiplet, and it is standard to be  identify it with 
 the real part $({\rm Re}\, T)$ of the multiplet. This identification  leads to the value (\ref{pred}) for $r$ which is in conflict with the BICEP2 data. 
 
One should stress that a different off-shell formulation (new minimal supergavity) \cite{CFPS} offers a different embedding of the Starobinsky
(higher curvature) model in
its dual standard supergravity. Here the two massive chiral multiplets of the (old minimal) formulation 
\cite{fv,Cecotti:1987sa}
are replaced by a massive vector multiplet whose (unique) scalar component is the inflaton \cite{FKLP,FKR}. Interestingly enough the massive vector multiplet model accomodates
most of the single field inflationary potentials once certain gobal conditions, related to the choice of gauging of the K\"ahlerian sigma models are fullfilled \cite{FFS}.
In particular the minimal supersymmetric extension of chaotic inflation 
\cite{chaotic,KLW}  is described by the Supersymmetric Stueckelberg model coupled to Supergravity, 
(corresponding to a flat K\"ahler space). This was first remarked in \cite{FKLP} and shown to be the zero-curvature limit \cite{KLR} of a continuous class of 
$SU(1,1)/U(1)$
gauged sigma models \cite{FKLP} called alpha attractors \cite{KLR}. This class includes for $\alpha=1$ the Starobinsky model itself. 
 
 However, it has been recently shown  in Ref. \cite{FKRim} that,  if one  identifies the inflaton not with 
 $({\rm Re}\, T)$ but rather with the imaginary $({\rm Im}\, T)$, one gets a  simple   chaotic inflationary model with quadratic potential \cite{chaotic}. This model is in perfect agreement with the BICEP2 data.
 
 This proposal  works as long as $({\rm Re}\, T)$  is stabilized at some finite value, which is not the case with the standard supersymmetric embedding of the Starobinsky model  \cite{KLr,jap}.
For this reasons one needs to modify the K\"ahler potential in order to stabilize the $({\rm Re}\, T)$, see Refs.   \cite{KLr,El} and \cite{FKRim}. 

A legitimate question is whether higher curvature corrections modify the agreement of the imaginary Starobinsky model and the BICEP2 data. Indeed, it  may happen that higher curvature terms spoil the form of the potential and lead to too low values of the tensor-to-scalar ratio $r$, making the model inappropriate  for inflation. We will see that 
the higher-order curvature terms make the structure of the imaginary Starobinsky model quite rich. Depending on the parameter range, they may not change the quadratic chaotic model or 
 may give rise to
scenarios where inflation is of the chaotic-like type with monomial potentials with fractional powers. For instance, 
the first correction in the Ricci scalar of the type $R^4$ may originate a potential with monomial power equal to $4/3$.

The paper is organized as follows. In section 2. we briefly summarize the imaginary Starobinsky model. In section 3 we discuss the impact of the $R^4$ terms and in section 4 we devote our attention to the  terms of the general form $R^{4+2N}$. Finally we conclude in section 5.

\section{The Imaginary Starobinsky Model}
In order to embed the   Starobinsky model in the old-minimal 
formaluation of the ${\cal N}=1$ supergravity, two chiral multiplets are needed
\cite{Cecotti:1987sa},
the inflaton multiplet $T$ and the goldstino multiplet $X$ \cite{ADFS}. 
The linear Goldstino multiplet $X$, at the level of linear representation, has not a well defined action since we can add an arbitrary $f(X)$ in the superpotential and an arbitrary $h(X,\bar X)$ in the Kahler potential $-3\log(T+\bar T +h)$. Also $h$ must be chosen to resolve the stabilization problem. To remove all  ambiguities on the $X$ self-interaction,  but
at  the same time to capture the goldstino feature,  we replace $X$ with its non linearly realized phase so that \cite{ADFS}
\be 
X^2=0. \label{X2}
\ee
In this way the sgoldstino field is no longer dynamical and all stabilization problems disappear. The goldstino multiplet \cite{FK,ADFS}  has been considered before for inflation in \cite{luis}.
Once this constraint is solved, it turns out that the lowest component of 
$X$ is proportional to  the goldstino bilinear $GG$ and, in particular,  $X$ is explicitly  written as 
\begin{equation}
X \ = \ \frac{GG}{2 \, F_X} \ + \ \sqrt{2}\, \theta\, G \ +\  \theta^2 F_X. \label{va1}
\end{equation}
The superfield $X$ can be used to elegantly embed the Starobinsky model into supergravity \cite{ADFS}. Indeed, the supergravity Lagrangian which describes the supersymmetric embedding of the Starobinksy model is 
\begin{align}
{\cal L} &= \int {\rm d}^2 \Theta\, 2 {\cal E} \Big{\{} \frac{3}{8} \, 
\Sigma \, e^{-K/3} + W  \Big{\}} + {\rm h.c.}, 
\label{sup1}
\end{align}
where 
$ \Sigma=\bar {\cal D}^2 - 8 {\cal R}$ is the chiral projection operator, 
and the K\"ahler potential and the superpotential  are (we set the reduced Planck scale to  unity)
\be
K = -3\, \text{ln}\,\big{(}{\cal T} + \bar{{\cal T}} - X \bar{X}\big{)}
\label{kahler}
\ee
and 
\be
W= 2 M \, X T-f X+\sigma X^2, \label{super}
\ee
respectively. 
Note that the chiral superfield $\sigma$ enforces the constraint (\ref{X2}). 
The theory (\ref{sup1}) can be written as a pure supergravity by noticing that the K\"ahler potential is of no-scale type for the $T$ multiplet \cite{no-scale}. Therefore, $T$ and $\overline{T}$ can be removed from the K\"ahler potential and appear in the superpotential which is now  written as
\be
W=  -6T {\cal R}+2 M \, X T-f X+\sigma X^2. 
\ee
Eliminating $T$ gives that 
\begin{eqnarray}
X=\frac{3{\cal R}}{M}, \label{XR}
\end{eqnarray}
and therefore the chiral supergravity multiplet ${\cal R}$  also satisfies
\be
{\cal R}^2=0, \label{R2}
\ee
due to the constraint imposed by $\sigma$.   Then, due to  (\ref{R2}), the  bosonic fields of the curvature chiral
scalar multiplet ${\cal R}$  appear only in its higher component, and the explicit  bosonic content  of  ${\cal R}$ is
\begin{eqnarray}
{\cal R}=\cdots+\theta^2 \mathscr{F}_{\cal R},
\end{eqnarray}
where 
\begin{eqnarray}
\mathscr{F}_{\cal R}=-\frac{1}{12} R - \frac{1}{18} A^m A_m + \frac{i}{6} {\cal D}^m A_m. 
\end{eqnarray}
By using Eq. (\ref{XR}), we may write the Lagrangian (\ref{sup1}) as
\be
{\cal L} = \Big{[} -3\frac{f}{M} + \frac{27}{M^2} {\cal R} \bar {\cal R} 
  \Big{]}_D,
\ee
the bosonic part of which is 
\be
e^{-1} {\cal L} = -3 \frac{f}{M}\, \mathscr{F}_{\cal R}  - 3 \frac{f}{M} \, \overline{\mathscr{F}} _{\cal R} 
+ \frac{27}{M^2} \mathscr{F}_{\cal R}   {\overline{\mathscr{F}}}_{\cal R}.
\label{f1}
\ee
After an appropriate rescaling of the metric, the Lagrangian (\ref{f1}) is explicitly written as
\begin{equation}
{\cal L} \ = \ \frac{1}{2}\ \left(R\, + \, \frac{2}{3}\ A_m^2\right)  \ + \ \frac{3}{16 M^2} \ \left(R\,+ \, \frac{2}{3}\ A_m^2\right)^2 \  +  \ \frac{3}{4M^2} \ ({\cal D}_m A^m)^2. \label{d15}
\end{equation}
It clearly describes an $(R+R^2)$ supergravity coupled to a pseudoscalar mode coming from ${\cal D}_m  A^m$.

On the other hand, the bosonic part of the original Lagrangian (\ref{sup1}) where the K\"ahler potential and superpotential are given by (\ref{kahler}) and (\ref{super}) respectively,  is explicitly  given by 
\be
e^{-1}{\cal L} =\frac{1}{2}  R  - \frac{3}{(T+\overline{T})^2} 
\partial_{\mu} T \partial^{\mu} \overline{T}-\frac{|MT-f|^2}{3(T+\overline{T})^2}. \label{st1}
\ee
Note that $X$ contributes to the scalar potential  since
\be 
F_X =e^{K}(K_{X\bar X})^{-1} W_X , 
\ee
and therefore it induces  a bosonic contribution although it contains no elementary scalar field.
Parametrizing $T$ as 
\begin{eqnarray}
T=e^{\sqrt{\frac{2}{3}}\phi}+i \sqrt{\frac{2}{3}}b,
\end{eqnarray}
and after putting the imaginary part of $T$  at its minimum $b=0$,
Eq. (\ref{st1}) is written as (with $\lambda=M^2/9,~f=M$)
\be
\label{OM7}
e^{-1}{\cal L}_1 = \frac{1}{2} R  -  \frac{1}{2} \partial_{\mu} \phi  \partial^{\mu} \phi 
-\ \frac{3}{4 } \,\lambda  \left(1-e^{-\sqrt{\frac{2}{3}} \phi}\right)^2.
\ee
This is  the standard Starobinsky model in the dual theory where the inflaton is the real part 
$({\rm Re}\, T)$ of the $T$ multiplet. However, as we have already mentioned, it cannot account for the amount of the gravitational waves 
reported by  BICEP2. 

The other possibility is to identify the inflaton not with $({\rm Re}\, T)$  but rather with the imaginary part $({\rm Im}\, T)$ . This idea has been proposed in \cite{FKRim}. In order for this to work, the $({\rm Re}\, T)$  needs be stabilized at some finite  value. This is not possible with the K\"ahler potential of Eq.~(\ref{kahler}) \cite{KLr,jap,El},
so a modification of the K\"ahler potential is needed \cite{KLr,El} (see also \cite{FKRim}). This
modification is however totally justified when thinking that inflation has to be followed by a period of reheating:  the chiral multiplet $T$ needs to be coupled to matter
\cite{FKRim}. 
Such couplings \cite{EKN}  change the K\"ahler potential according to 

\be
K=-3 \ln\left(T+\overline{T}-X\overline{X}+ (T+\oT)^n F(\Phi_i)+{\rm h.c.}\right)+ K_m(\Phi_i,\overline{\Phi}_i).
\ee
If  
$\langle  D_iW\rangle=0$,  all matter scalars are stabilized at 
$\langle\Phi_i\rangle$. If we now assume that $F(\langle \Phi_i\rangle)=m$, 
the effective  K\"ahler potential  for  the $T$ and $X$ multiplets is given by
\be
K=-3 \ln\left(T+\oT-X\overline{X}+ m (T+\oT)^n\right). \label{kah}
\ee
Other  modification of the K\"ahler potential are also possible \cite{KLr,El,olivelimit}. 
Then for a superpotential of the form
\begin{eqnarray}
W(T,X)=W(T,X,\langle\Phi_i\rangle)=  3 \sqrt{\lambda} \, X (T -f) , \label{potmod}
\end{eqnarray}
the resulting potential  turns out to be
\begin{eqnarray}
 V_T=3\lambda \frac{|T-f|^2}{\left[T+\bar{T}+m (T+\overline{T})^n\right]^2}, \label{VF2m}
 \end{eqnarray}
whereas the K\"ahler metric is 
\begin{eqnarray}
K_{T\overline{T}}=3\frac{1+mn(T+\overline{T})^{n-2}\Big{[}(3-n)(T+\overline{T})+m(T+\overline{T})^n\Big{]}}{\left[T+\overline{T}+m (T+\overline{T})^n\right]^2}. \label{kahmet}
\end{eqnarray}
In terms of the real and imaginary parts of $T$, we may express the potential as
\begin{eqnarray}
\label{p}
V(\phi,b)=\frac{3}{4}\lambda \frac{\Big{(}1-f\,e^{-\gamma\phi}\Big{)}^2}{\Big{(}1+2^{n-1}m\,
e^{(n-1)\gamma\phi}\Big{)}^2}+\frac{3}{4}\gamma^2\lambda \frac{e^{-2\gamma\phi}}{\Big{(}1+2^{n-1}m
e^{(n-1)\gamma\phi}\Big{)}^2}b^2,  \label{pot}
\end{eqnarray}
where $\gamma=\sqrt{2/3}$. The bosonic Lagrangian may be written as
\begin{eqnarray}
{\cal L}=\frac{1}{2} R-\frac{1+mn (2e^{\gamma\phi})^{n-2}\big{[}
2(3-n)e^{\gamma\phi}+m (2e^{\gamma\phi})^n\big{]}}{2\big{(}1+m (2e^{\gamma\phi})^{n-1}\big{)}^2}\Big{(}\partial_\mu\phi\partial^\mu\phi+e^{-2\gamma\phi}\partial_\mu b\partial^\mu b\Big{)}-V(\phi,b). \label{ll0}
\end{eqnarray}
Let us note that when $m=0$, the 
scalars parametrize the K\"ahler space  $SU(1,1)/U(1)$ and (\ref{ll0})    reduces to  (\ref{OM7}) for $b=0$. 
For a generic value of $m$, the scalar manifold is deformed such that only a $U(1)$ isometry is preserved. In this case, as $m$ tends towards $m=-(2f)^{1-n}$,
the minimum of the potential in the field $\phi$ gets steeper and steeper when $n$ goes to unity. This behaviour is similar with that of the model considered in \cite{KLr}. However, in order to  simplify the discussion here, let us take the particular value \cite{FKRim}
\begin{eqnarray}
\label{point}
 m=-n^{-1}(2f)^{1-n}, 
 \end{eqnarray} 
 with $n\neq 1$, for which 
there is a minimum $\phi=\phi_0=\ln f^{1/\gamma}$ independently of the   value of $b$. 
At the minimum
  the potential for the imaginary part of the $T$-field turns out to be
\begin{eqnarray}
V_{\rm eff}(b)=V(\phi_0,b)=\frac{ n^2\lambda}{2f^2(n-1)^2}\,b^2. \label{v01}
\end{eqnarray}
Since, 
\begin{eqnarray}
K_{T\oT}\Big{|}_{\phi_0}=\frac{3n}{4f^2} ,
\end{eqnarray}
upon redefining $\chi=b\sqrt{n}/f$  
the Lagrangian  with a canonically normalized kinetic term for the $\chi$ field is explicitly written as
\begin{eqnarray}
{\cal{L}}=\frac{1}{2} R-\frac{1}{2}\partial_\mu \chi\partial^\mu \chi-
\frac{1}{2}m_\chi^2\chi^2,
\end{eqnarray}
where 
\begin{eqnarray}
m_\chi^2=\frac{n\lambda}{(n-1)^2}.
\end{eqnarray}
 Thus the imaginary Starobinsky model coupled to matter becomes  just the 
 minimal chaotic inflation with quadratic potential and  it  predicts 
\begin{eqnarray}
n_S-1\approx-\frac{2}{N}=-0.04\left(\frac{50}{N}\right), ~~~r\approx \frac{8}{N}=0.16\left(\frac{50}{N}\right), 
\end{eqnarray}
which is in  good agreement with the BICEP2 data \cite{FKRim}.

 We should stress at this point that the imaginary Starobinsky model coupled to matter, 
 although it is inspired by the $(R+R^2)$ gravity,  it differs  from that. This is however expected because one needs to couple it to matter in order to reheat the universe after inflation. In this sense, the so-called ``Starobinsky limit" \cite{olivelimit} does not really exist.

\section{$(R+R^2+R^4)$ extension of the imaginary Starobinsky model} 
As we wrote in the introduction, it is a legitimate to ask what happens to the conclusions of the previous section when higher-order terms in the Ricci scalar are added. In particular, $R^4$ terms can be included as 
\begin{eqnarray}
{\cal L} = \left[ -3 + \frac{3}{ \lambda} {\cal R} \bar {\cal R} 
+\xi_4
{\cal R}\bar {\cal R}
\bar \Sigma({\cal R})    \Sigma(\bar {\cal R})
 \right]_D. \label{rNew00}
\end{eqnarray}
This is of the form discussed in Ref. \cite{Cecotti:1987sa}. In particular, (\ref{rNew00}) can be written as 
\begin{align}
{\cal L} =& -3\left[ T+\bar T-X \bar X -\frac{z}{12} X\bar X B\bar B
 \right]_D+3\sqrt{\lambda}\Big{[}X(T-f)\Big{]}_F 
 +\Big{[}\sigma X^2\Big{]}_F \nonumber \\
 &+\frac{3}{2\sqrt{\lambda}}\left[Q(B-\Sigma(\bar {\cal R})
 \right]_F+{\rm h.c.} \label{rNew2}
\end{align}
where $Q,B$ are chiral superfields and $z=4\lambda \xi_4$. This is standard supergravity with 
K\"ahler potential 
\begin{eqnarray}
K=-3 \ln \left(T+\bar T-X \bar X -Q \bar X-\bar Q X-\frac{z}{12} X\bar X B\bar B\right), \label{kaplu0}
\end{eqnarray}
and superpotential
\begin{eqnarray}
W=3\sqrt{\lambda}\, X (T-f )+\sigma X^2 +\frac{3}{2\sqrt{\lambda}}Q B.
\end{eqnarray}
 The observation made in Ref. \cite{Cecotti:1987sa} was that the K\"ahler potential (\ref{kaplu0}) is not  plurisubharmonic due to the presence of the $Q\bar X+\bar Q X= Y_+\bar Y_+-Y_-\bar Y _-$ term,  where  
$Y_\pm=(Q\pm X)/\sqrt{2}$. This leads then to a ghost state corresponding to the negative eigenvalue of the K\"ahler metric $K_{i\bar{j}}$. However, since in our case we have $X^2=0$, it is easy to see that the only propagating scalar is the $T$-field with the standard kinetic term $K_{T\bar T} \partial T\partial \bar T=3 \partial T\partial \bar T/(T+\bar T)^2$ and there is no any ghost state. In fact the  fields $X,Q,B$  contribute in the bosonic sector  only in the scalar potential.
In order to determine the latter, the  non-zero components of the K\"ahler metric are needed. For $X|=0$ they  turn out to be  
\be
&&K_{T\oT}=\frac{3 \left(1-m (n-3) n t^{n-1}+m^2 n t^{2( n-1)}\right)}{t^2 \left(1+m
   t^{n-1}\right)^2},  ~~~ K_{T\overline{X}}=-\frac{3 Q \left(1+m n t^{n-1}\right)}{t^2 \left(1+m t^{n-1}\right)^2}, \\
   &&
 K_{X\oT}=-\frac{3 \overline{Q} \left(1+m n t^{n-1}\right)}{t^2 \left(1+m t^{n-1}\right)^2}, ~~~K_{X\overline{X}}= \frac{3
   \left(Q\overline{Q}+\left(m t^n+t\right) (1+\frac{z}{12}\, B \overline{B} )\right)}{t^2\left(1+m
   t^{n-1}\right)^2}, \\
   && K_{X\overline{Q}}=\frac{3}{t(1+m t^{n-1})} ,~~~K_{Q\overline{X}}= \frac{3}{t(1+m t^{n-1})},
\ee
where $t=(T+\oT)$ and the interaction $m |T+\bar T|^n$ has been incuded in the K\"ahler potential. The potential then takes the form
 \begin{align}
V&=-e^{K/3}\left(
K_{T\oT}F_T\overline{F} _{\oT} + K_{T\overline{X}}F_T\overline{F}_{\overline{X}}+K_{X\oT}F_X\overline{F}_{\oT}+K_{X\overline{X}}F_X{\overline{F}_{\overline{X}}}+K_{X\overline{Q}}F_X \overline{F}_{\overline{Q}}+K_{Q\overline{X}}F_Q\overline{F}_{\overline{X}}\right)
\nonumber \\
&+e^{2K/3}(F_TD_TW+F_X D_XW+F_QD_QW+F_BD_BW+h.c.)-3 e^K W  \bar{W}.
\end{align}
Then,  the fields $F_B,{\overline{F}}_{\overline{B}}$, 
$F_T,\overline{F}_{\oT}$, $F_Q,\overline{F}_{\overline{Q}}$, $Q,\overline{Q}$ and $B,\oB$ can be eliminated by their equation of motions, the solution of which is  
\be 
&&\overline{B}=2\sqrt{\lambda} F_X\, , ~~~
B=2\sqrt{\lambda}\overline{F}_{\overline{X}}, ~~~F_Q=4\lambda z\, F_X^2 \oF_{\oX}, ~~~\oF_{\overline{Q}}=4\lambda z\, F_X \oF_{\oX}^2  \nonumber \\
&&Q=\overline{Q}=F_T=\overline{F}_{\oT}=F_B=\oF_{\oB}=0. 
\ee
Using the expressions above for the auxiliaries, the potential $V$ turns out to be

\begin{eqnarray}
V=-\frac{1}{[T+\overline T+m (T+\overline T)^n]^2}
\Big{\{}-3\lambda^{1/2}\big{[}F_X(T-f)+\overline{F}_X (\overline{T}-f)\big{]}+3 F_X\overline F _X+z (F_X\overline F _X)^2\Big{\}}. \label{vq}
\end{eqnarray}

In order to determine $F_X,\oF_{\oX}$, 
let us write  $V$ in Eq. (\ref{vq}) as 
\begin{eqnarray}
V={\cal A}F_X+\bar {\cal A}\oF_{\oX}+{\cal B} F_X\overline F _{\oX}+
{\cal S} (F_X\overline F _{\oX})^{N+2}
\end{eqnarray}
where 
\begin{eqnarray}
&&{\cal A}=\frac{3\lambda^{1/2}(T-f)}{[T+\overline T+m (T+\overline T)^n]^2},\nonumber \\
&&{\cal B}=-\frac{3}{[T+\overline T+m (T+\overline T)^n]^2}, \nonumber \\
&&{\cal S}=-\frac{z}{[T+\overline T+m (T+\overline T)^n]^2}\, . \label{ABS}
\end{eqnarray}
Then  the equations of motion for $F_X,\oF_{\oX}$ are
 
\be
0={\cal A} + {\cal B} \overline{F}_X + 2 {\cal S}   F_X \overline{F}_X^2, \ \ 
0=\overline{{\cal A}} + {\cal B} F_X + 2 {\cal S}   \overline{F}_X F_X^2,
\ee
These two equations can be combined into the single equation  
\be
\alpha = Y (1 + \beta Y )^2,  \label{quartic}
\ee
where
\be
\alpha &=& \frac{{\cal A} \overline{{\cal A}}}{{\cal B}^2},
\\
\beta &=& \frac{2 {\cal S}}{{\cal B}},
\\
Y &=& F_X \overline{F}_X.
\ee
The solution to the  equation (\ref{quartic}) above can be expressed as 
\be
Y = \frac{2}{3 \beta} \big{(}\cosh s -1\big{)},
\ee
with
\be
s = \frac{1}{3 } \arccosh\Big{(}\frac{27}{2} \alpha \beta +1\Big{)}.
\ee
The full  scalar potential has then  the following compact form
\be
V =- {\cal B} Y - 3 {\cal S} Y^2 . \label{potF}
\ee
By employing  the K\"ahler potential (\ref{kah}) and the superpotential (\ref{potmod}), 
we get 
  the full potential 
\begin{align}
V(\phi,b)&=
\frac{3 \cosh ^2\Big{[}\frac{1}{3} 
\cosh ^{-1}\left(1+9 z \lambda \left[b^2 \gamma^2+\left(e^{\gamma  \phi }-f\right)^2\right]\right)\Big{]}}{4ze^{2\gamma\phi} \left(1+m  \left(2 e^{\gamma  \phi }\right)^{n-1}\right)^2}-\nonumber \\
&-\frac{3 \cosh
   \Big{[}\frac{1}{3} 
\cosh ^{-1}\left(1+9 z \lambda \left[b^2 \gamma^2+\left(e^{\gamma  \phi }-f\right)^2\right]\right)\Big{]}}{4 ze^{2\gamma\phi} \left(1+m  \left(2 e^{\gamma  \phi }\right)^{n-1}\right)^2}.\label{pot}
\end{align}
where $\gamma=\sqrt{2/3}$. The K\"ahler metric is still given by Eq. (\ref{kahmet}) and therefore  the bosonic Lagrangian turns out to be
\begin{eqnarray}
{\cal L}=\frac{1}{2} R-\frac{1+mn (2e^{\gamma\phi})^{n-2}\big{[}
2(3-n)e^{\gamma\phi}+m (2e^{\gamma\phi})^n\big{]}}{2\big{(}1+m (2e^{\gamma\phi})^{n-1}\big{)}^2}\Big{(}\partial_\mu\phi\partial^\mu\phi+e^{-2\gamma\phi}\partial_\mu b\partial^\mu b\Big{)}-V(\phi,b). \label{ll}
\end{eqnarray}
For  the particular value  $
 m=-n^{-1}(2f)^{1-n}$ adopted here (the value $n=1$ is excluded as  the K\"ahler potential (\ref{kah}) does not depend on $T$ in this case), there is still the minimum at
\be 
\phi_0=\ln f^{1/\gamma},\label{fo}
\ee
 independently of the   value of $b$ and the field $\phi$ is stabilized there and does not participate to the
 dynamics. Since again $
K_{T\oT}\Big{|}_{\phi_0}=(3n/4f^2)$ ,
one should use once more  the canonically normalized field  $\chi=b\sqrt{n}/f$.  
 At the minimum, 
   the potential for the imaginary part of the $T$-field turns out to be
   
\begin{eqnarray}
\label{aaa}
V(\phi_0,b)&=&
\frac{3 n^2}{2 f^2 (n-1)^2 \, z}
\cosh \left[\frac{1}{3} \cosh ^{-1}\left(1+6z\lambda b^2\right)\right] \sinh ^2\left[\frac{1}{6} \cosh ^{-1}\left(1+6z\lambda b^2
   \right)\right].
%
\end{eqnarray}
Let us discuss the various limits. Since we are interested in large field models of inflation to produce a sizable amount of tensor modes,  consider first the case  $\lambda z=4 \lambda^2\xi_4\lll 1$ and the 
 range  
 
\be
  1\ll b^2\lll(6\lambda z)^{-1}.
  \ee
The condition $\lambda^3\xi_4\lll 1$ implies that the higher-order curvature term $R^4$ is subdominant
over the $R^2$ piece. 
In such a case the potential (\ref{aaa}) takes the form
\begin{eqnarray}
V(\phi_0,b)=\frac{  n^2\lambda }{2 f^2  (n-1)^2}b^2-\frac{ n^2 \lambda^2 z}{9 f^2 
   (n-1)^2}b^4+{\cal O}\left(z^{3/2}\right). \label{ap}
\end{eqnarray}
The first term is just (\ref{v01}) whereas the second term is the first small correction due to $R^4$ correction.  Therefore,
the mass term is dominating and we recover, as we should, the imaginary Starobinsky model.

However, there exists another interesting limit, namely $\lambda z=4 \lambda^2\xi_4\ggg 1$ and

\be
\label{aa}
b^2\gg (6\lambda z)^{-1}, 
\ee
in which, along all the inflationary trajectory, the potential turn out to be to leading order

\be
V_{\rm eff}(b)=
  \frac{3(12\lambda)^{2/3}   n^2}{16 f^2  (n-1)^2
   z^{1/3}}\,  b^{4/3}. \label{poteff}
\ee
The corresponding Lagrangian  for the  canonically normalized kinetic term for large values of $b$ is written as 
\begin{eqnarray}
{\cal{L}}=\frac{1}{2} R-\frac{1}{2}\partial_\mu \chi\partial^\mu \chi
-g_{_0} \, \chi^{4/3}, \label{in10}
\end{eqnarray}
where 
\begin{eqnarray}
   g_0=\frac{3(12\lambda)^{2/3}   n^{4/3}}{16 f^{2/3}  (n-1)^2
   z^{1/3}}. 
\end{eqnarray}
In such a limit the imaginary Starobinsky model gives a monomial chaotic model of inflation with fractional power. 
The corresponding spactral index and the scalar-to-tensor ratio is then
\begin{eqnarray}
n_S-1=-\frac{5}{3N}=0.96\left(\frac{50}{N}\right), ~~~r=\frac{16}{3N}=0.1\left(\frac{50}{N}\right),
\end{eqnarray}
which is still compatible with the BICEP2 data.
Notice that for  $\lambda z=\lambda^3\xi_4\lll 1$ one can also consider the limit $b^2\gg (6\lambda z)^{-1}$. In this case the Lagrangian 
is of the form (\ref{in10}), but the observationally interesting dynamics happens when the quadratic potential dominates, that is when $b^2$ becomes smaller than $(6\lambda z)^{-1}$. Of course, one can consider the fine-tuned option that 
$(\lambda z)^{-1}$ takes any value  between unity and ${\cal O}(10^2)$. In such a case the largest observationally interesting scales to  exit the Hubble radius will experience the $b^{4/3}$ potential and the smallest the $b^2$ potential.

 It should be also noted   that the Lagrangian (\ref{in10}) could be derived easily  even without knowing the full potential  (\ref{pot}). Indeed, for large values of 
$Y$ satisfying $Y\gg1/\beta$, Eq. (\ref{quartic}) is written as 
\begin{eqnarray}
 \alpha\approx \beta^2 Y^3,
 \end{eqnarray} 
 which specifies $Y$ to be given by
 \begin{eqnarray}
 Y\approx \left(\frac{\alpha}{\beta^2}\right)^{1/3}=\left(\frac{9 \lambda }{4 z^2}\,|T-f|^2\right)^{1/3}.
 \end{eqnarray}
 Then, it is straightforward to verify that the potential turns out to be (in this large $Y$ region)
\begin{eqnarray}
V(\phi,b)\approx
\label{eq:1} 
\frac{3 \lambda^{1/3}e^{-2\gamma\phi}}{4z^{2/3}\Big{(}1+2^{n-1}m\,
e^{(n-1)\gamma\phi}\Big{)}^2}\left(
\, \Big{[}b^2+(e^{\gamma \phi}-f)^2\gamma^{-2}\Big{]}^{1/3}+
\frac{3\gamma}{2}(\lambda z)^{1/3}  \Big{[}b^2+(e^{\gamma \phi}-f)^2\gamma^{-2}\Big{]}^{2/3}\right). \nonumber\\
&&
\end{eqnarray}
 Interestingly, 
 Eq. (\ref{eq:1}) has a minimum again at 
 $\phi_0=\ln f^{1/\gamma}$. Therefore it reduces to the form (\ref{poteff}) once 
 $\phi$ is stabilized at its minimum. In addition, the condition $Y\gg 1/\beta$  
leads to  
\begin{eqnarray}
\chi^2\gg \frac{9}{4}\frac{n}{f^2\lambda z}, \label{lower}
\end{eqnarray}
which  is consistent with the condition (\ref{aa})
 used to derive Eq.   (\ref{poteff}). 
 

\section{$(R+R^{4+2N})$ extension of the imaginary Starobinsky model}
Similarly,  we may consider the  general theories $(R+R^2+R^{4+2N})$. 
%
Such theories are described by 

\be
{\cal L} = \left[ -3 + \frac{3 }{\lambda} {\cal R} \bar {\cal R} 
+\frac{\xi_N}{16^{N}} 
{\cal R} \bar {\cal R}
\Big{(}\Sigma(\bar {\cal R})\bar \Sigma({\cal R})\Big{)}^{N+1} 
%
\right]_D, \label{rN}
\ee
the bosonic part of which  turns out to be (with ${\cal R}^2=0$)
\be
e^{-1} {\cal L} = -3 
\mathscr{F}_{\cal R}  - 3 
\overline{\mathscr{F}} _{\cal R} 
+ \frac{3}{\lambda}\mathscr{F}_{\cal R}   \overline{\mathscr{F}} _{\cal R}  
+ 16  \xi_N \left(  \mathscr{F}_{\cal R}   \overline{ \mathscr{F}} _{\cal R}  \right)^{N+2}. \label{rNN}
\ee
After an appropriate rescaling of the metric, the Lagrangian in (\ref{rNN}) is explicitly written as  
\begin{align}
{\cal L} & =  \frac{1}{2}\ \left(R\, + \, \frac{2}{3}\ A_m^2\right)  \ + \ \frac{1}{48 \lambda} \ \left(R\,+ \, \frac{2}{3}\ A_m^2\right)^2 \  +  \ \frac{1}{48 \lambda} \ ({\cal D}_m A^m)^2\nonumber \\
&+ \frac{\xi_4}{9\cdot 12^{2N+2}}
\Big{[} \left(R\,+ \, \frac{2}{3}\ A_m^2\right)^2+4({\cal D}_m A^m)^2\Big{]}^{2+N} \label{d17}
\end{align}
and it is clearly an $(R+R^2+R^{4+2N})$ supergravity. It is interesting to note that for $N=-2$, Eq. (\ref{d17}) describes an $(R+R^2)$ theory with a cosmological constant, which can easily also checked from the superspace form  of the action 
in Eq. (\ref{rN}).  
%
The latter  can equivalently be written as 
\begin{align}
{\cal L} =& -3\left[ T+\bar T-X \bar X -
\frac{z}{12\cdot 4^N} X\bar X B^{N+1}\bar B ^{N+1}
 \right]_D+3\sqrt{\lambda}\Big{[}X(T-f)\Big{]}_F 
 +\Big{[}\sigma X^2\Big{]}_F \nonumber \\
 &+\frac{3}{2\sqrt{\lambda}}\left[Q(B-
 \Sigma(\bar{\cal R})\right]_F+h.c. \label{rNew2}
\end{align}
where all constraints have been taken into account and $z=4^{N+1}\lambda \xi_N$ here.
 This is again standard supergravity with 
K\"ahler potential 
\begin{eqnarray}
K=-3 \ln \Big{(}T+\bar T-X \bar X -Q \bar X-\bar Q X-
\frac{z}{12\cdot 4^N} X\bar X B^{N+1}\bar B ^{N+1}\Big{)}, \label{kaplu}
\end{eqnarray}
and superpotential
\begin{eqnarray}
W=3\sqrt{\lambda}\, X (T-f )+\sigma X^2 +\frac{3}{2\sqrt{\lambda}}Q B.
\end{eqnarray}
 Here again, although the K\"ahler potential (\ref{kaplu}) is not a plurisubharmonic due to the presence of the $Q\bar X+\bar Q X$ term,  there are no ghost due to the nilpotency of $X$ as in the $N=0$ case discussed in the previous section.      

In order to determine the potential, the non-zero components of the K\"ahler potential at $X|=0$ are needed. By using (\ref{kaplu}) we find 

\be
&&K_{T\oT}=\frac{3 \left(1-m (n-3) n t^{n-1}+m^2 n t^{2( n-1)}\right)}{t^2 \left(1+m
   t^{n-1}\right)^2},  ~~~ K_{T\overline{X}}=-\frac{3 Q \left(1+m n t^{n-1}\right)}{t^2 \left(1+m t^{n-1}\right)^2}, \\
   &&
 K_{X\oT}=-\frac{3 \overline{Q} \left(1+m n t^{n-1}\right)}{t^2 \left(1+m t^{n-1}\right)^2}, ~~~K_{X\overline{X}}= \frac{3
   \left(Q\overline{Q}+\left(m t^n+t\right) (1+\frac{z}{12\cdot 4^N}\, B^{N+1} \bar B^{N+1} )\right)}{t^2\left(1+m
   t^{n-1}\right)^2}, \\
   && K_{X\overline{Q}}=\frac{3}{t(1+m t^{n-1})} ,~~~K_{Q\overline{X}}= \frac{3}{t(1+m t^{n-1})},
\ee
where the interaction $m|T+\bar T|^n$ has been included. The potential is then
 
 \begin{align}
V&=-e^{K/3}\left(
K_{T\oT}F_T\overline{F} _{\oT} + K_{T\overline{X}}F_T\overline{F}_{\overline{X}}+K_{X\oT}F_X\overline{F}_{\oT}+K_{X\overline{X}}F_X{\overline{F}_{\overline{X}}}+K_{X\overline{Q}}F_X \overline{F}_{\overline{Q}}+K_{Q\overline{X}}F_Q\overline{F}_{\overline{X}}\right)
\nonumber \\
&+e^{2K/3}(F_TD_TW+F_X D_XW+F_QD_QW+F_BD_BW+h.c.)-3 e^K W  \bar{W}, 
\end{align}
and the solution of the equations of motion for the fields $F_B,{\overline{F}}_{\overline{B}}$, 
$F_T,\overline{F}_{\oT}$, $F_Q,\overline{F}_{\overline{Q}}$, $Q,\overline{Q}$ and $B,\oB$ gives 

\be 
&&\overline{B}=2\sqrt{\lambda} F_X\, , ~~~
B=2\sqrt{\lambda}\overline{F}_{\overline{X}}, ~~~F_Q=(N+1)(4\lambda)^{N+1} z\, F_X^{N+2} \oF_{\oX}^{N+1},\nonumber \\
&&~~~\oF_{\overline{Q}}=(N+1)(4\lambda)^{N+1} z\, F_X^{N+1} \oF_{\oX}^{N+2}, ~~~Q=\overline{Q}=F_T=\overline{F}_{\oT}=F_B=\oF_{\oB}=0. 
\ee
Using the expressions above for the auxiliaries, the potential $V$ turns out to be

\begin{eqnarray}
V=-\frac{1}{[T+\overline T+m (T+\overline T)^n]^2}
\Big{\{}-3\lambda^{1/2}\big{[}F_X(T-f)+\overline{F}_X (\overline{T}-f)\big{]}+3 F_X\overline F _X+z (F_X\overline F _{\oX})^{N+2}\Big{\}}. \label{po0}
\end{eqnarray}
In order to integrate out the auxiliaries $F_X,\oF_{\oX}$, let us write $V$ in Eq. (\ref{po0}) as 
\begin{eqnarray}
V={\cal A} F_X+\bar {\cal A}\oF_{\oX}+{\cal B} F_X\overline F _{\oX}+
{\cal S} (F_X\overline F _{\oX})^{N+2}
\end{eqnarray}
where  ${\cal A},~{\cal B} $ and ${\cal S}$ are given in Eq. (\ref{ABS}).
The equations of motion for $F_X$ are 
\be
&&0={\cal A} + {\cal B} \overline{F}_X + (2+N) {\cal S}_N  \overline{ F} _X  (F_X \overline{F}_X)^{N+1},\nonumber \\
&&0=\overline{{\cal A}} + {\cal B} F_X + (2+N) {\cal S}_N  F_X  (F_X \overline{F}_X)^{N+1},
\ee
which can be reduced to the single  equation  

\be
\alpha_N = Y (1 + \beta_N Y^{N+1} )^2,  \label{power}
\ee
where
\be
&&\alpha_N = \frac{{\cal A} \overline{{\cal A}}}{{\cal B}_N^2}
=\lambda |T-f|^2,\nonumber \\
&&\beta_N = \frac{(N+2) {\cal S}}{{\cal B}}=(N+2)\frac{z}{3},\nonumber \\
&&Y=F_X\oF_{\oX}
\ee
The potential turns out then to be
\be
V=-{\cal B} Y -(2N+ 3) {\cal S} Y^{N+2}. \label{potF}
\ee
 Here,  Eq. (\ref{power}) cannot be solved exactly for general $N$. 
In the limit $Y\ll 1/\beta_N^{1/(N+1)}$, we find that 
\begin{align}
V&\approx -a_N {\cal B}-(2N+3) {\cal S}_N a_N^{N+2}=\nonumber \\
&=\frac{3\lambda|T-f|^2}{[T+\overline T+m (T+\overline T)^n]^2}\, \Big{\{}1-\frac{\lambda^{N+1}\,  z}{3}\,  |T-f|^{2N+2}+\ldots\Big{\}}.
\end{align}
For $m=-(2f)^{1-n}/2n$ there is a minimum at $\phi_0=\ln f^{1/\gamma}$ where the potential is written as 

\begin{eqnarray}
V(\phi_0,b)=\frac{  n^2\lambda }{2 f^2  (n-1)^2}b^2-\frac{2^N\lambda^{N+2}\, z\, n^2 }{3^{2+N} f^2 
   (n-1)^2}\, b^{2N+4}+\cdots \label{VF2m}
\end{eqnarray}
for the imaginary part of the $T$ field, which coincides with Eq. (\ref{ap}) for $N=0$. The first term is the potential (\ref{VF2m}) and the second terms is the first correction due to the $R^{4+2N}$ term. As the first term always dominates in the present approximation, we recover the quadratic  imaginary Starobinsky model.

Let us solve Eq. (\ref{power}) in the opposite limit $Y\gg1/\beta_N^{1/(N+1)}$. In this case, Eq. (\ref{power}) is solved by
\begin{eqnarray}
 Y\approx \left(\frac{\alpha_N}{\beta_N^2}\right)^{1/(2N+3)}=
  \left(\frac{9 \lambda }{(2+N)^2 z^2}\,|T-f|^2\right)^{1/(2N+3)}
 \end{eqnarray}
and the potential turns out to be
\begin{align}
 V(\phi,b)&\approx -(2N+ 3) {\cal S}_N Y^{N+2}=
\frac{(2p-1)^{2p-1}\,e^{-2\gamma\phi}\, (9 \gamma^2 \lambda)^p}{4 p^{2p}\Big{(}1+2^{n-1}m\,
e^{(n-1)\gamma\phi}\Big{)}^2 z^{2p-1}} \Big{[}b^2+(e^{\gamma \phi}-f)^2\gamma^{-2}\Big{]}^{p}, \label{vio}
\end{align}
where 
\be 
p=(N+2)/(3+2N).
\ee 
The minimum of the potential is still at $\phi_0=\log f^{1/\gamma}$ so that the potential for the imaginary part of the $T$ field when its real part is stabilized at $\phi_0$ turns out to be
\be
V^{(N)}_{\rm eff}(b)=
\frac{(2p\!-\!1)^{2p-1}\,n^2\,  (6 \lambda)^p}{4 p^{2p} f^2 (n\!-\!1)^2 z^{2p-1}} \, b^{2p}.
\ee
The Lagrangian for the canonically normalized field 
$\chi=b\sqrt{n}/f$ is written as 
\begin{eqnarray}
e^{-1}{\cal{L}}_N=\frac{1}{2} R-\frac{1}{2}\partial_\mu \chi\partial^\mu \chi
-g_{_N} \, \chi^{2p}, \label{in0}
\end{eqnarray}
where 
\begin{eqnarray}
g_{_N}=\frac{(2p\!-\!1)^{2p-1}\,n^{2-p}\,  (6\lambda)^p}{4p^{2p} f^{2(1-p)} (n\!-\!1)^2 z^{2p-1}}.
\end{eqnarray}
To determine the possible values of $p$, let us note that the potential (\ref{vio}) is positive for $z(2N+3)>0$. Therefore,  
we have $N>-3/2$ for $z>0$ and $2p$ is a ractional number 
with  range  
\be 
1<2p\leq 2.
\ee
In fact, the possible values for $2p$ are $2p=(2,4/3,5/6,8/7,\cdots)$ for $N=(-1,0,1,2,\cdots)$.
In this range, the imaginary Starobinsky model predicts a monomial chaotic model of inflation with fractional powers between 1 and 2.  Note that the  value 
$2p=2$  corresponds to $N=-1$, {\it i.e.} to  $R+R^2$ theory.

The condition $Y\gg(1/\beta_N)^{1/(N+1)}$ we have used to derive Eq. (\ref{in0}) is explicitly written as 
\begin{eqnarray}
 \chi^2\gg \frac{3}{2}\left(\frac{pz}{3(2p\!-\!1)}\right)^{\frac{2p-1}{p-1}}\frac{n}{\lambda f^2}.
 \label{min}
\end{eqnarray}
If $z^{\frac{2p-1}{p-1}}/\lambda\lll1$, then the potential $\chi^{2p}$ dominates along all the inflationary trajectory and one finds that 
the spectral index and the scalar-too-tensor ratio of (\ref{in0}) are
\begin{eqnarray}
n_S-1=-\frac{p+1}{N}<0.04 \left(\frac{50}{N}\right)\, , ~~~r=\frac{8p}{N}<0.16 \left(\frac{50}{N}\right),
\end{eqnarray}
where we have made use of the condition $p<1$. In this case we see that the predictions are consistent with the observations.

We should point out that one may advocate general theories of the form

\be
{\cal L} = \left[ -3 + \frac{3}{ \lambda} {\cal R} \bar {\cal R} 
+\sum_N \frac{\xi_N}{16^N} 
{\cal R} \bar {\cal R}
\Big{(}\Sigma(\bar {\cal R})\bar \Sigma({\cal R})\Big{)}^{N+1} 
%
\right]_D, \label{rNnn}
\ee
In this case, the exact scalar potential cannot explicitly be found as it is not possible to completeley integrate out all auxiliaries. However, the analysis above is still valid for the dominant term in the above sum.

\section{Conclusions}
The recent  BICEP2 data on the B-mode polarization of the CMB anisotropies oblige inflationary model-builders to reconsider their preferred models. In Ref. \cite{FKRim} it has been recently shown that the Starobinsky model, which seemed to be preferred by the Planck data and subsequently disfavored by the BICEP2 data, is in fact in full agreement with the latter if, once embedded in supergravity, one identifies the inflaton field with the imaginary part of the chiral multiplet in the dual formulation. The resulting potential turns out to be that of the simple quadratic chaotic inflation, which
predicts a large amount of tensor modes. 
In this short paper we have shown that, depending on the parameter space,  the  higher-order corrections in the curvature term can either preserve the above conclusion or change  it giving  rise for large values of the field
to monomial potentials whose power is fractional and between 2 and 1. The original quadratic piece
may or may not  dominate according to the parameter range, rendering the structure of the imaginary Starobinsky model coupled to matter quite rich.

\section*{Acknowledgements}
S.F.  would like to acknowledge R. Kallosh, A. Linde, and M. Porrati for enlightening conversations and critical remarks on the interpretation of the gravitational axion as the inflaton of chaotic inflation. A.K. thanks F. Farakos for correspondence. S.F. is supported by ERC Advanced Investigator Grant n. 226455,
``Supersymmetry, Quantum Gravity and Gauge Fields (Superfelds)". A.R. is supported by the Swiss National
Science Foundation (SNSF), project ``The non-Gaussian Universe" (project number: 200021140236).
This research was implemented under the “ARISTEIA” Action of the “Operational Programme 
Education and Lifelong Learning” and is co-funded by the European Social Fund
(ESF) and National Resources. 
This work is partially supported by European Union's Seventh
Framework Programme (FP7/2007-2013) under REA grant agreement n. 329083.

\end{document}